\documentclass[twocolumn,showpacs,showkeys,preprintnumbers,
               amsmath,amssymb,floatfix,10pt]{revtex4}
\usepackage{graphicx}
\usepackage{txfonts}
\usepackage[small]{titlesec}
\titlespacing*{\section}{0pt}{0.4cm}{0.1cm}
\titlespacing*{\subsection}{0pt}{0.4cm}{0.1cm}
\begin{document} 
   \title{Mutational dynamics of influenza A viruses:\\
          a principal component analysis\\
          of hemagglutinin sequences of subtype H1}

   \author{Yves-Henri Sanejouand}

   \affiliation{UFIP, UMR 6286 of CNRS,\\Facult\'e des Sciences et des Techniques,
    Nantes, France.\\
    Yves-Henri.Sanejouand@univ-nantes.fr}
   \date{October 4, 2017}
   \maketitle

\section*{Abstract}

A principal component analysis of a multiple sequence alignement
of hemagglutinin sequences of subtype H1 has been performed,
the sequences being encoded 
using the amino-acid property that maximizes the weight
of the major component.
In the case of this alignment, it happens to be a well-known
hydrophobicity scale.
 Interestingly, sequences coming from human have large positive amplitudes
along the major component before 2009, and large negative ones afterwards.
This means that the 2009 pandemic was associated to
a major change in the hydrophobicity pattern of hemagglutinin.

The present analysis also highlights the high variability of
viral sequences coming from swine.  
At a more general level, the method proposed in this paper
allows to describe a sequence coming from an alignment with a set of numbers,
the original point being that the choice of the corresponding property is driven by the data.
This approach should allow the application of numerous methods to the study of large multiple sequence alignments. 
 
\section*{Introduction}

Because it does not require any assumption about the underlying 
population genetic model, and also because it allows to study 
large datasets at a negligible computational cost, 
principal component analysis (PCA)~\cite{Ringner:08}
has been used for long for analyzing multiple sequence 
alignments (MSA)~\cite{Vanheel:91,Valencia:95,Almeida:03,Jalview,Weigt:13}.

To this end, since PCA deals with numerical quantities, 
each sequence symbol needs to be associated to a set of numbers. 
In the case of nucleic acids, an obvious choice is a binary code~\cite{Valencia:95,Jalview}
where, for instance, $\{1,0,0,0\}$ corresponds to adenine,
$\{0,1,0,0\}$ to cytosine, \textit{etc}. 

For proteins, because there are twenty common amino-acid residues,
doing so would yield extremely sparse matrices, since many residues
are never observed at a given position, even in the
case of large alignments, like the one considered in the present study.
This issue has for instance been addressed by using instead
frequencies of amino-acid residues, in whole genomes~\cite{Suhre:03},
or counts of 
pairs of residues found in each considered sequence~\cite{Vanheel:91}.
In the present study, it is addressed by associating a single numerical property 
to each residue, the property being chosen among the 544 properties 
gathered in the amino acid index database~\cite{Kawashima:00,Kawashima:08}, 
so that the relative weight of the major component of the PCA 
is the largest.
 
As a first application, this approach is used for analyzing the MSA of 
influenza A hemagglutinin sequences belonging to subtype H1.
Gaining a better understanding of the mutational dynamics of this subtype
may indeed prove of particular importance, since it has been involved in
several pandemics, noteworthy the 1918-1919 one~\cite{Taubenberger:99}, which killed
at least 50 million people~\cite{Mueller:02},
but also in the latest one, in 2009-2010~\cite{Rambaut:09,Kawaoka:09}.

\section*{Methods}

\subsection*{Multiple sequence alignment}

17688 hemagglutinin (HA) sequences of subtype H1 were 
retrieved~\footnote{On September 6$^{th}$, 2016.}
from the NCBI influenza virus resource~\cite{Ivr:08}, 
sequences coming from laboratory viral strains being disregarded,
as well as redundant ones. 
Since obtaining an accurate MSA of a large number of sequences
can prove challenging~\cite{Poch:11,Higgins:13,Notredame:14},
and because H1 sequences have high levels of sequence identities, 
being at least 75\% identical to each other~\cite{Sanejouand:17},
like in a previous study~\cite{Sanejouand:17},
pairwise alignments were performed, with BLAST~\cite{Blast:97} 
version 2.2.19, taking as query the long H1 sequence of 
virus A/Thailand/CU-MV10/2010 (genbank accession number HM752477). 
MVIEW~\cite{Mview:98}, version 1.60.1, was then used for converting the BLAST output into an actual MSA.

Including gaps, this MSA is 575 residues long.
For performing PCA, 205 sites were considered, those 
with little variability being disregarded, namely,
all sites where the same amino-acid residue is found in at least
99\% of the sequences,
as well as those that are not observed in crystal structure 4EEF~\cite{Baker:12},
the latter being mostly at both ends of the 
MSA~\footnote{There are 498 amino-acid residues in each HA monomer of 4EEF.}.

The 4EEF structure was used for illustrative purposes,
as well as for residue numbering which is, 
like in most available crystal structures~\footnote{For instance, ten X-ray structures
of the 1918 HA have been determined and the H3 numbering was used for nine of them.},  
the H3 numbering, even though 4EEF is a structure of the 1918 HA,
from strain A/South Carolina/1/1918~\cite{Taubenberger:99}
(genbank accession number AF117241.1).

Though the rate of evolution of H1 sequences over the last century
has not been spectacular, contemporary sequences being on average
more than 80\% identical to the 1918-1919 sequences~\cite{Sanejouand:17}, 
note that, due to the large number of H1 sequences taken into account,
the variability of the 205 retained sites is high,  
$\approx$10 different residues being observed at each site of the MSA, on average.

\begin{figure*}
\centering
\includegraphics[width=17 cm]{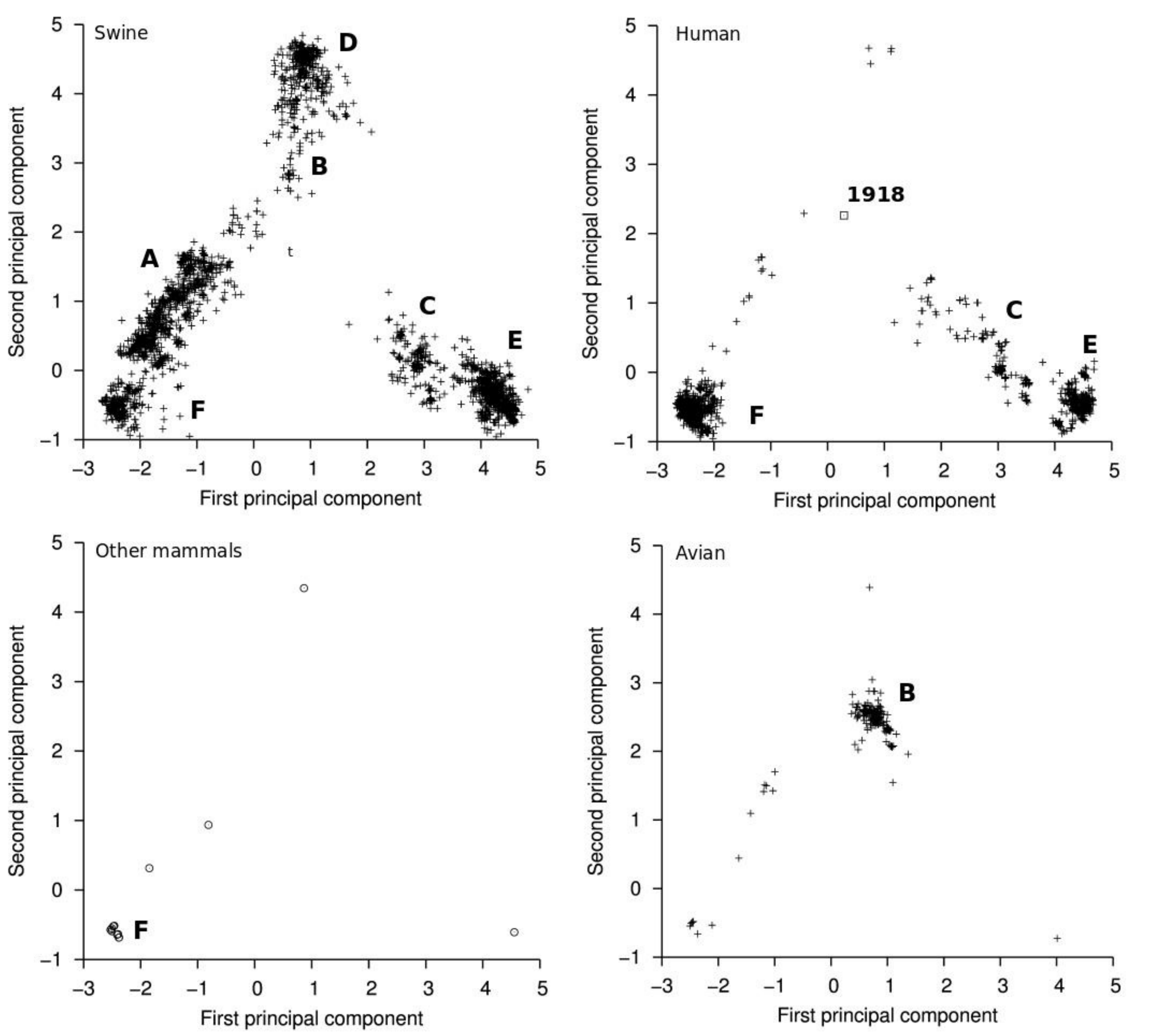}
\vskip -0.3 cm
\caption{\textit{Projections of HA sequences on the two first PCA components}. 
Sequences come from swine (top left), human (top right), birds (bottom right)
or from mammals other than swine and human (bottom left). 
Main sequence clusters are labelled \textbf{A-F}.  
Open square: the 1918 sequence.
}
\label{Fig:proj}
\end{figure*}

\subsection*{Principal component analysis}

Let us associate a set of $n$ numerical properties to a given residue $i$,
$p_{i,1}, \dots, p_{i,n}$, so that a sequence $k$ of length $N$ can be described as a 
vector $\mathbf{s}_k=\{ p_{1,1}, \dots, p_{N,n} \}$ of dimension $d=n N$.
A MSA can then be described as a matrix: 
\[
\mathbf{S} = 
\begin{pmatrix}
\mathbf{s}_1 - \mathbf{s}_{ref} \\
\vdots \\
\mathbf{s}_k - \mathbf{s}_{ref} \\
\vdots \\
\mathbf{s}_m - \mathbf{s}_{ref} \\
\end{pmatrix}
\]
where $\mathbf{s}_{ref}$
is a reference sequence and $m$, the number of sequences.
$\mathbf{C}$, the covariance matrix, 
of dimension $d \times d$, 
is:
\[
\mathbf{C} = \frac{1}{m} \mathbf{S}^ T \mathbf{S} 
\]
$\mathbf{A}$, the orthogonal matrix with the principal components, and $\mathbf{\Lambda}$,
the diagonal matrix with their weights, are obtained
by diagonalizing $\mathbf{C}$~\cite{Rao:64}:
\[
\mathbf{A}^T \mathbf{C} \mathbf{A} = \mathbf{\Lambda}
\]
Note that the weight of a principal component gives the proportion of
the variance of the sequences, with respect to the reference one, that
is captured by the component.

On the other hand, since the principal components form a basis set,  
sequence $k$ can be described as a set of amplitudes (projections)
along the principal components:
\begin{equation}
q_i = \mathbf{a}_i \cdot ( \mathbf{s}_k - \mathbf{s}_{ref} )
\label{eq:proj}
\end{equation}
where $q_i$ is the
amplitude of sequence $k$ along component $i$, 
$\mathbf{a}_i=\{ a_{1,i}, \dots, a_{d,i} \}$
being the $i^{th}$ component, that is, the $i^{th}$ eigenvector of $\mathbf{C}$,
and $a_{j,i}$ the coefficient of component $i$ for the $j^{th}$ property
of sequence $k$~\footnote{When $n=1$, this is the coefficient of the component for residue $j$.}. 

Hereafter, the $n=1$ case is considered and the reference sequence,
for which $\mathbf{q} = \mathbf{0}$, is the average sequence of the MSA. 
Gaps and unknown residues are treated as follows:  
the property value of a gap is assumed to be the average value
at the considered site; the property value of an unknown residue
is assumed to be the value obtained for the closest sequence having a known
residue at that site. 

Projections (eqn~\ref{eq:proj}) were only performed for long enough sequences, 
namely, for the 11869 sequences 
where a known amino-acid residue is found in at least 90\% of the 205 selected
sites of the MSA.   
 
\section*{Results}

\subsection*{Choice of the amino-acid property}

All 544 properties of the amino acid index database~\cite{Kawashima:00,Kawashima:08}
were tried one after the other, a PCA of the MSA with the 17688 H1 sequences
being performed for each of them. The weight of the major (first) component
varies between 39.9\% and 63.9\% of the overall variance (the trace of $\mathbf{C}$)
of the sequence dataset. Interestingly, properties yielding the largest 
weight for the major component
are well known hydrophobicity scales. Indeed, 
the scales that are, according to our criterion, the three best ones 
were built with 
residue contact matrices~\cite{Vendruscolo:05},
mean polarities~\cite{Wolfenden:88} 
and amino-acid 
partition energies~\cite{Miyazawa:99}.

Being the best one, the former was retained for
further analysis.   
It corresponds to the following residue ranking: EKRSDQGNPHTAMWYCFLVI.
As expected for an hydrophobicity scale, the two basic (RD) and the two acidic (ED) residues
are at one end of the scale, namely, among the five first ones,
while the four last ones (FLVI) are the residues that are the most often considered
to be the most hydrophobic ones~\cite{Trinquier:98}.

The weights of the second and third components are 9.8\% and 3.5\%, respectively. 
Thus, nearly three quarters (74\%) of the fluctuations of the 17688 H1 sequences
can be described with two components \textit{only} (among 205), 
most remaining ones being of little significance. 
As a matter of fact, only eight components have a weight of more than 1\%.

\subsection*{Projections on the two first components}

Sequence fluctuations are, by definition~\cite{Rao:64}, the largest along the 
major component. As shown in Figure~\ref{Fig:proj},
where the projections of H1 sequences on the two major
components (eqn~\ref{eq:proj}) are plotted, 
most human sequences (Figure~\ref{Fig:proj}, top right) belong
to a pair of clusters, coined \textbf{E} and \textbf{F}, which correspond to extreme
values of the amplitude along the major component: $q_1 \approx 4.5$
and $q_1 \approx -2.5$, respectively. 
Both clusters
are also observed with swine sequences (Figure~\ref{Fig:proj}, top left), 
while most sequences from mammals
other than swine and human (Figure~\ref{Fig:proj}, 
bottom left) belong to cluster \textbf{F}. 
Note that this latter point is likely to be a consequence of 
the lack of data for these species before 2009.  
Indeed, the single sequence found in cluster \textbf{E},
from a giant anteater, was obtained in 2007
while the only other sequence obtained before 2009, from a ferret,
belongs to yet another one, coined \textbf{A}.
A sequence belonging to cluster \textbf{D} was also found in 2013, coming from a wild boar.

Most avian sequences (93\% of them) belong to 
a fourth cluster (Figure~\ref{Fig:proj}, bottom right), coined \textbf{B}.
Since complete avian sequences are known 
since 1979~\footnote{Two 1917 avian sequences were determined, but they are partial ones.},
this result confirms that a strong evolutionary pressure is at work
in avian species~\citep{Sanejouand:17}, which limits the variability of avian H1 sequences. 

Interestingly, the 1918 sequence colocalizes with cluster \textbf{B} 
(Figure~\ref{Fig:proj}, top right),
further supporting the hypothesis of an avian origin for the 1918-1919 
pandemic~\citep{Taubenberger:99}. However, seven sequences coming from swine
with collection dates between 1931 and 1942 are also located close
to the 1918 sequence. Since these latter sequences are also the closest ones
in terms of sequence identity~\citep{Sanejouand:17}, based on our sole analyses
of the hemagglutinin sequences, the hypothesis 
that the 1918 virus actually came from swine would be more likely. 

The limited variability of avian sequences helps 
highlighting a key result of the present analysis, 
namely, the spectacular variability of sequences coming from swine (Figure~\ref{Fig:proj}, top left).
On the one hand, swine sequences are found in all major clusters
observed with sequences of other species. On the other hand, two clusters
(\textbf{A} and \textbf{D}) are mostly populated by
sequences coming from swine.

\begin{figure}
\centering
\includegraphics[width=8.5 cm]{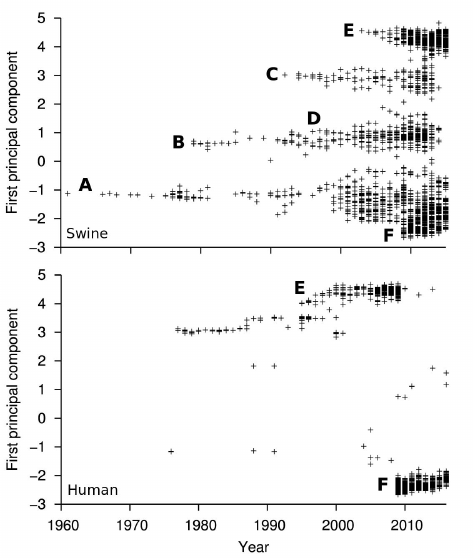}
\vskip -0.3 cm
\caption{\textit{Projections of HA sequences on the major PCA component, as a function of time}. 
Sequences come from swine (top) or human (bottom).
The year correspond to the collection date of each sequence, 
as provided by the NCBI influenza virus ressource. 
Main sequence clusters are labelled \textbf{A-F},
like in Figure~\ref{Fig:proj}.  
Following the amplitude of the second component (not shown) allows 
to pinpoint when cluster \textbf{D} popped up.
}
\label{Fig:projtime}
\end{figure}

\subsection*{Projection as a function of time}

Figure~\ref{Fig:projtime} shows the evolution of the projections of the sequences
on the major component, as a function of their collection date. For sequences
coming from swine, this analysis highlights two striking features: first, 
a new cluster of swine sequences has been popping up every five-ten years 
(lately: \textbf{E} in 2003, \textbf{F} in 2009). Second, 
half of them seem to have vanished after 2014 (clusters \textbf{B-D}).

For sequences coming from human, 
our analysis highlights the fact that, a given year, almost all of them belong to a given
cluster, with a switch from cluster \textbf{E} to cluster \textbf{F} occurring in 2009.  
Indeed, before the 2009 pandemic, no sequence belonging to cluster \textbf{F} was found 
while, after 2009, sequences belonging to cluster \textbf{E} are rare (see Figure~\ref{Fig:projtime}). On the other hand, the fact that 
the cluster the closest to cluster \textbf{F} is
cluster \textbf{A} (see Figure~\ref{Fig:proj}) suggests that the former derives
from the later, that is, since most sequences of cluster \textbf{A} come from swine,
it supports the hypothesis that the 2009 pandemic has its origin in this species~\citep{Kawaoka:09}.  

Sequences with a collection date before 1960 are rare. As a consequence,
following their projections on the first component (not shown)
does not allow to check if, for instance, sequences coming 
from human have experienced other jumps from a sequence cluster
to another, like the 2009 one. This seems however likely since, while for the 1918
sequence $q_1 \approx 0.2$, it was significantly higher in the thirties ($q_1$ in the 1.5--2 range).

\begin{figure}
\centering
\includegraphics[width=9.0 cm]{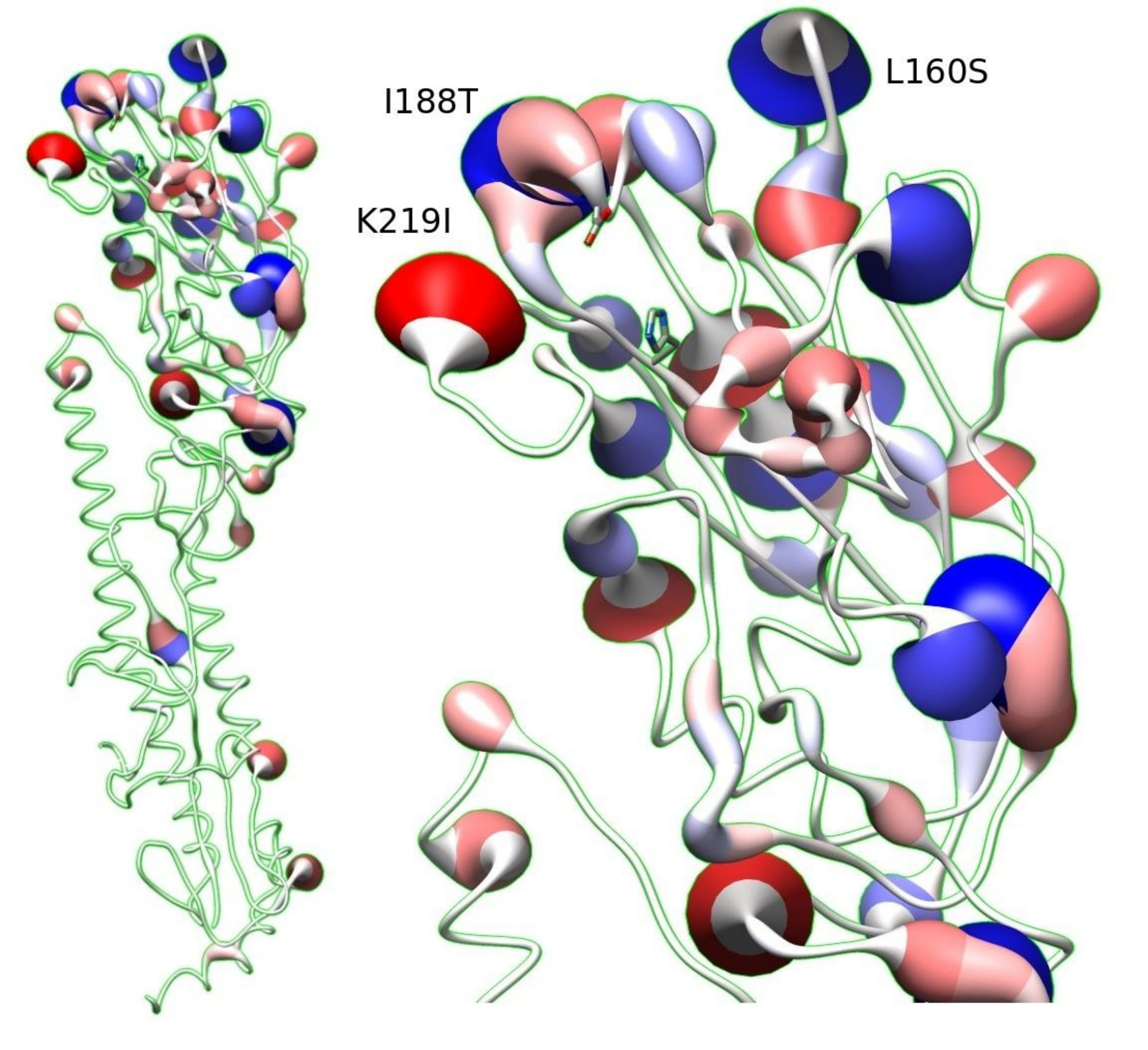}
\vskip -0.3 cm
\caption{\textit{Change in the hydrophobicity pattern of hemagglutinin}. 
The width of the worm is proportional to the absolute value of the 
coefficient of the major PCA component for the residue. The colour gives the
sign of the coefficient. Red means that the residue was polar before 2009
and has been hydrophobic since then. Blue means the opposite.
The three residues 
with large absolute coefficients that are the closest to the receptor binding site are labelled,
the residue the most often observed in sequences of human origin before 2009 being mentioned first, the residue the most often observed since then
being mentioned last.
Drawn with UCSF Chimera (version 1.11.2)~\citep{Chimera}.
}
\label{Fig:major}
\end{figure}

\subsection*{Analysis of the major component}

Figure~\ref{Fig:major} shows that the coefficients 
of the major component are much larger (whatever their sign)
on the head of hemagglutinin (residues 53-269), where the binding site of the receptor stands.
Indeed, on the rest of hemagglutinin (86 analyzed sites) the absolute value of the
coefficient is always less than 0.16
while, on the head of hemagglutinin, it is larger 
for 14 residues, raising up to 0.25~\footnote{Being an eigenvector, a component is normalized, that is, the sum of the square of its coefficients is one.}. 

Moreover, 
six of these residues have positive coefficients, namely,
A103I, T155V, A169I, S203F, K219I, N269I,
the coefficients being negative for the other eight ones, namely,
L53K, L78S, I80S, V133N, L160S, I188T, V205G, I244T,
the residue given first being the most commonly found one before 2009 in sequences of human origin, while the second is the most commonly found afterwards~\footnote{H3 numbering.}. 
This means that, though the overall hydrophobicity of the head of hemagglutinin
has not changed significantly in 2009, the hydrophobicity pattern there has changed dramatically.

I188T, the residue with the third largest coefficient (in terms of absolute values),
was a glycine in the 1918 HA sequence. 
Interestingly, at variance with all the other residues of the 1918 sequence,
Gly 188 has not been observed
again in H1 sequences of human origin~\citep{Sanejouand:17}. 
This suggests that mutations at this position may play a key role in the development of pandemics. It further calls for a dedicated monitoring of such mutations.

\section*{Conclusion}

Encoding the hemagglutinin sequences belonging to
subtype H1 with the hydrophobicity of their residues,
using a well known scale~\cite{Vendruscolo:05},
allows to describe $\approx$64\% of the fluctuations of these sequences
with a single principal component, which
corresponds to a major change in the pattern of hydrophobicity 
on the head of hemagglutinin (Fig.~\ref{Fig:major}), where the receptor binding site stands.
This change occurred in 2009 (Fig.~\ref{Fig:projtime}), suggesting
that it is involved in the pandemic, probably by modifying extensively the antigenicity
of hemagglutinin, thus helping the virus to escape recognition by the immune system.

Taken together, the two major components allow to delineate several clusters of sequences
(Fig.~\ref{Fig:proj}), highlighting the reduced variability of sequences
of avian origin, most of them being included in a single cluster,
in contrast with sequences from swine, which are found in in at least six different ones.

Projecting the swine sequences on the major component as a function of time (Fig.~\ref{Fig:projtime}) shows that, while new clusters appear regularly, namely,
every five-ten years, several seem to have vanished after 2014. 
As a consequence, most actual sequences from swine belong to the same two clusters
where sequences of human origin are found. 

In the case of hemagglutinin sequences, describing sequences with a single property per residue proved enough for getting meaningful components. It is likely that for other alignments using more properties per residue could prove helpful.  

\clearpage


\end{document}